\documentclass[UKenglish,a4paper]{lipics-v2019-nosc}
\title{Plane-filling trails}
\author{Herman Haverkort}{Universität Bonn, Germany \and \url{http://herman.haverkort.net/} }{cs.herman@haverkort.net}{}{}
\authorrunning{H. J. Haverkort}
\Copyright{Herman J. Haverkort}
\relatedversion{A short version of this paper with supplements has been conditionally accepted to the 2020 Computational Geometry Week Media Exposition.}
\supplement{For software and additional figures, visit \url{http://spacefillingcurves.net}.}
\EventEditors{Sergio Cabello and Danny Z. Chen}
\EventNoEds{2}
\EventLongTitle{36th International Symposium on Computational Geometry (SoCG 2020)}
\EventShortTitle{SoCG 2020}
\EventAcronym{SoCG}
\EventYear{2020}
\EventDate{June 23--26, 2020}
\EventLocation{Z\"{u}rich, Switzerland}
\EventLogo{socg-logo}
\SeriesVolume{164}
\hideLIPIcs
\nolinenumbers
\newcommand\pftrail{\textsf{pftrail}}
\newcommand\collada{\textsc{collada}}

\begin{document}

\maketitle

\begin{abstract}
The order in which plane-filling curves visit points in the plane can be exploited to design efficient algorithms. Typically, the curves are useful because they preserve locality: points that are close to each other along the curve tend to be close to each other in the plane, and vice versa. However, sketches of plane-filling curves do not show this well: they are hard to read on different levels of detail and it is hard to see how far apart points are along the curve. This paper presents a software tool to produce compelling visualisations that may give more insight in the structure of the curves.
\end{abstract}

\subparagraph{Plane-filling curves}
A plane-filling curve is a continuous surjective mapping $f$ from the unit interval to a subset of the plane that has positive area, that is, Jordan content. Although such a mapping cannot be one-to-one, an unambiguous inverse can be defined with a tie-breaking rule. Thus, the mapping provides an order in which to process points in the plane. Famous examples include Pólya's triangle-filling curve~\cite{Pol1913} and square-filling curves by Peano~\cite{Pea1890} and Hilbert~\cite{Hil1891}. Continuity of the mapping is not always required: if we drop this requirement, we speak of plane-filling \emph{traversals}. Z-Order~\cite{Mor66} is an example that is often applied in practice. 

Plane-filling traversals and their inverses have been used to design efficient solutions for various applications, including indexing of points in the plane, geometric algorithms and data structures, finite element methods, load balancing in parallel computing, improving cache utilization in computations on large matrices or images, combinatorial optimization, image compression, information visualization, and sonification~\cite{Bad13,Hav17}. It is therefore interesting to see the differences between the various plane-filling traversals that have been proposed.

\begin{figure}
\centering\includegraphics{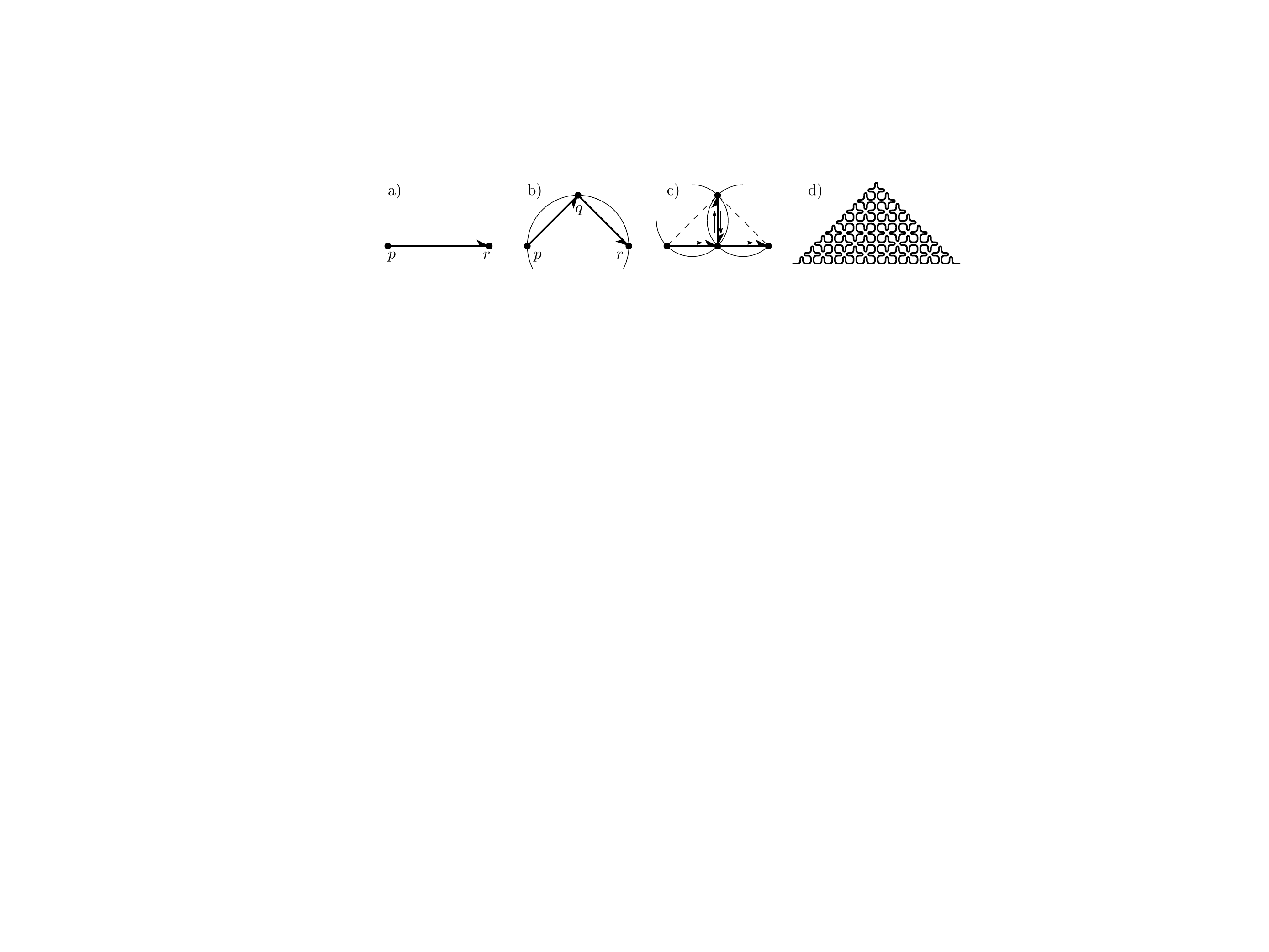}
\caption{Pólya's triangle-filling curve.}
\label{fig:polyadefinition}
\end{figure}

\subparagraph{Defining a plane-filling curve}
Plane-filling traversals are usually visualised in a way that follows their definition. Consider Pólya's curve. To define it, we start with a single line segment (Figure~\ref{fig:polyadefinition}a). We refine this simple drawing as follows. Let $p$ and $r$ be the end points of the original line segment. Imagine a circle with centre line $pr$ and draw another point $q$ halfway on the circle as we follow it clockwise from $p$ to $r$. Erase the original line segment $pr$ and replace it by two smaller segments $pq$ and $qr$ (Figure~\ref{fig:polyadefinition}b). Next, refine the drawing again by applying the same refinement procedure to each segment, but this time changing the orientation: to find the new intermediate points, we now follow the circles in counterclockwise direction. To indicate this change in orientation, we add an arrow head to $pr$ on the left side, and put the arrow heads for $pq$ and $qr$ on the right side. Thus, two line segments become four segments (Figure~\ref{fig:polyadefinition}c). Note that the middle two segments lie on top of each other, but they have different directions. If we repeat this refinement process six more times, alternating clockwise and counterclockwise, and move all points slightly so that the curve does not back up on itself, we get Figure~\ref{fig:polyadefinition}d. If we continue ad infinitum, the curve fills a right isosceles triangle. 

\begin{figure}
\centering\includegraphics{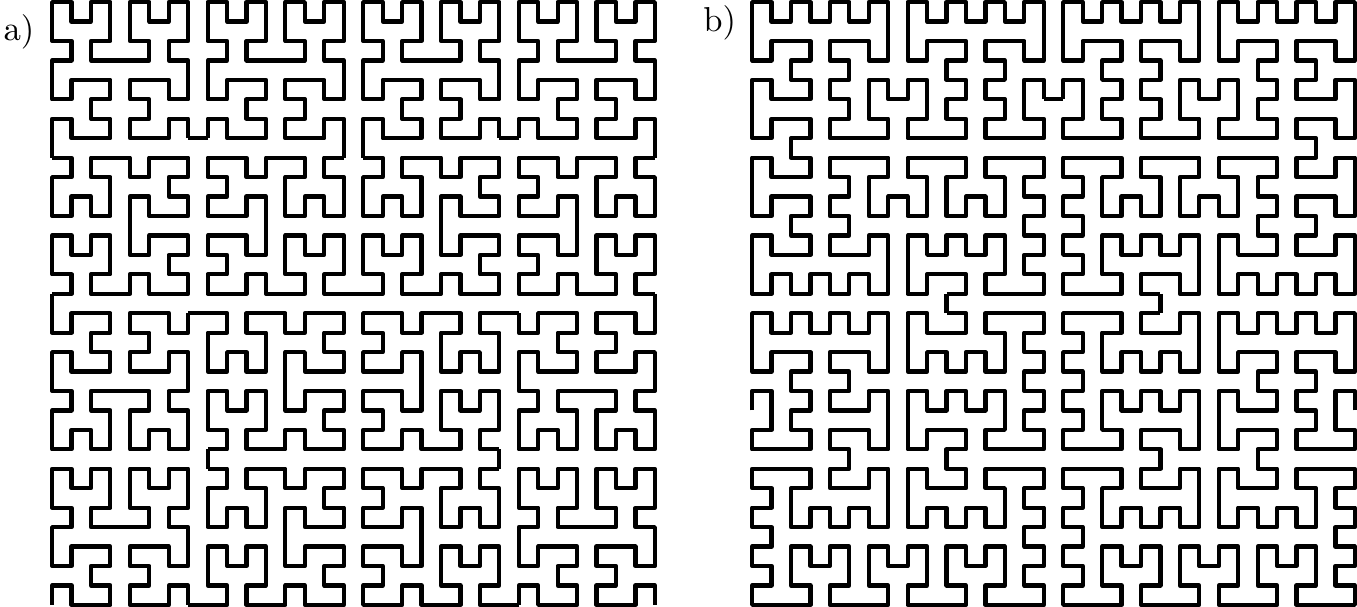}
\caption{Sketches of a) the Hilbert curve~\cite{Hil1891} and b) an $\Omega$ section of the $\beta\Omega$-curve~\cite{Wie02}.}
\label{fig:hilbertsketch}
\end{figure}

\begin{figure}
\centering
\includegraphics{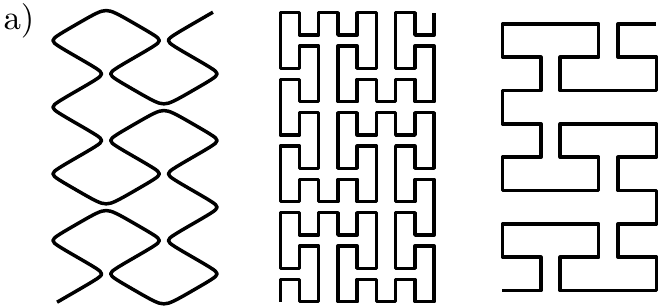}\quad\quad\quad
\includegraphics{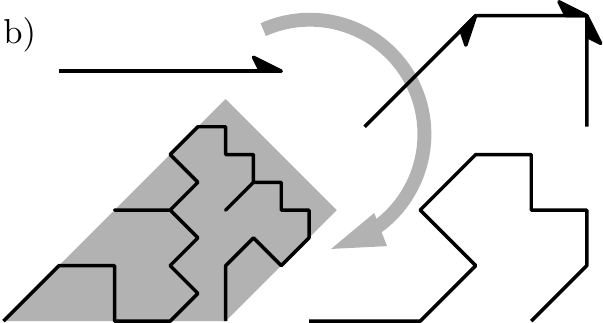}
\caption{a) Three traditional-style sketches of Peano's curve, mapped to a $\sqrt{3} :1$ rectangle. The left sketch connects the end points of the defining segments; the other two sketches connect their centre points, on different levels of refinement. Note how these different sketches give conflicting impressions of the curve, in particular of the main axes of movement. 
b) A sketch of a trapezoid-filling curve.}
\label{fig:confusingsketches}
\end{figure}

\subparagraph{The challenge of visualising plane-filling curves}
Figures \ref{fig:polyadefinition}b and \ref{fig:polyadefinition}d are typical of the way in which plane-filling curves are usually sketched. Neither figure makes it clear in an instant in what order the curve fills what parts of the plane---not to mention showing the curve's locality-preserving properties and violations thereof. Try comparing, for example, Hilbert's curve in Figure~\ref{fig:hilbertsketch}a to the $\beta\Omega$-curve in Figure~\ref{fig:hilbertsketch}b (a promising alternative~\cite{YL06}). 
Given a pair of points in the plane, these drawings do not allow us to see at a glance how long the path from one point to another along the curve is, and what regions of the plane are visited on the way.
Furthermore, the impression one gets of the curve depends heavily on how one chooses to define it and on the details of how it is sketched. Figure~\ref{fig:confusingsketches}a shows three sketches that all sketch the same curve, and Figure~\ref{fig:confusingsketches}b shows a sketch of a trapezoid-filling curve that is nothing else than the first three quarters of Pólya's curve: none of this is visually obvious from the drawings.

\begin{figure}
\centering\hbox to\hsize{%
a) \vtop{\hbox{}\vskip-3ex\hbox{\includegraphics[width=2.5in]{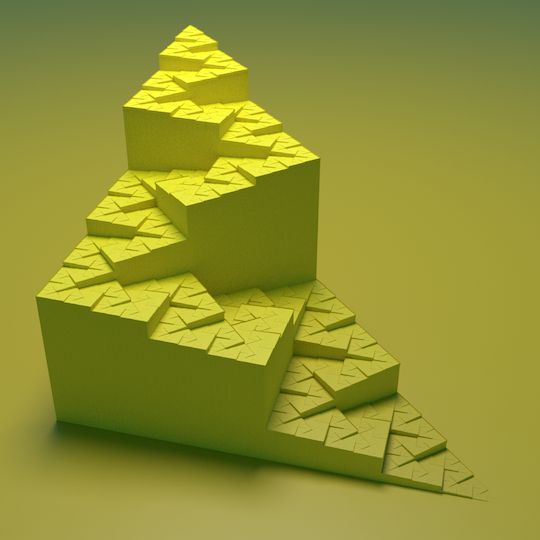}}}\hfill
b) \vtop{\hbox{}\vskip-3ex\hbox{\includegraphics[width=2.5in]{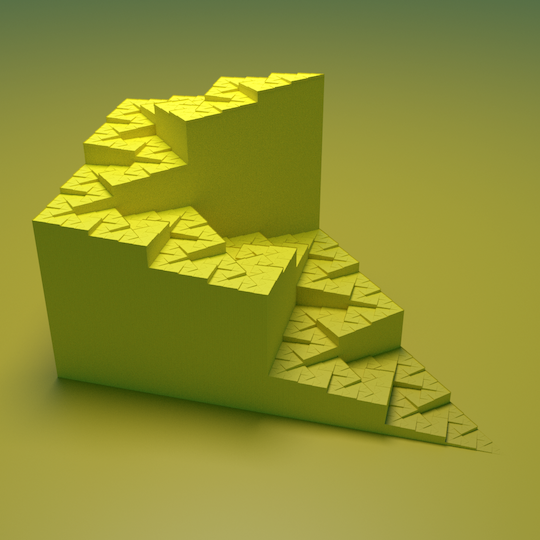}}}}
\caption{a) Pólya's curve. b) The trapezoid-filling curve sketched in Figure~\ref{fig:confusingsketches}b.}
\label{fig:polyatrail}
\end{figure}

\subparagraph{Visualisation as three-dimensional landscapes}
To visualise plane-filling curves and traversals more clearly, I developed a tool \pftrail. The tool reads a definition of a plane-filling curve and produces a \emph{plane-filling trail}, a model of the curve on a three-dimensional landscape, in which each point $f(t) = (x,y)$ of the curve is rendered as a point $(x,y,t)$. Thus the curve becomes a steadily ascending path in the landscape, see Figure~\ref{fig:polyatrail}. At a low resolution, the concept can be seen in action in a Hilbert curve marble run design by Ortiz~\cite{Ort18}. At higher resolutions, we obtain a clear visualisation of the locality-preserving and locality-violating properties of the curve that can be studied at different levels of detail. High, steep slopes reveal pairs of points that are close in the plane but far apart along the curve. Narrow corridors reveal sections between points that are relatively close to each other along the curve, but far apart in the plane. Wide corridors show sections of the curve that have good locality-preserving properties in both directions. The global course of the curve is easy to follow, but the image also facilitates studying the curve in more detail. Moreover, the visualisation is independent of what definition of the curve is used, out of multiple equivalent definitions. For example, the fact that the trapezoid-filling curve is simply the first three quarters of the Pólya curve is now obvious, see Figure~\ref{fig:polyatrail}. The visualisation gives the user the possibility to study the curve without any bias towards an arbitrary underlying tessellation.

\begin{figure}
\hbox to\hsize{%
\includegraphics[width=0.49\hsize]{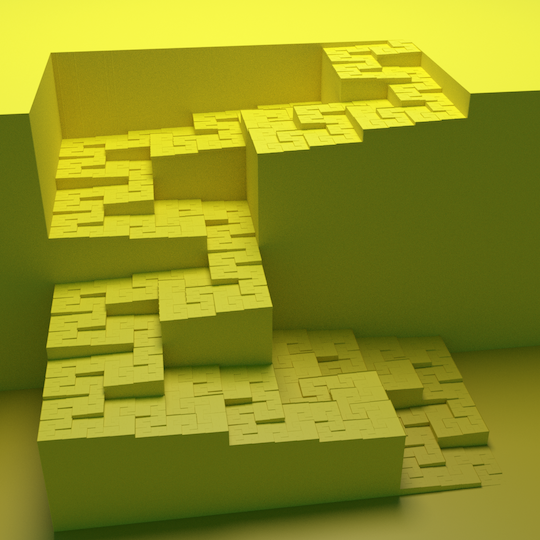}\hfill
\includegraphics[width=0.49\hsize]{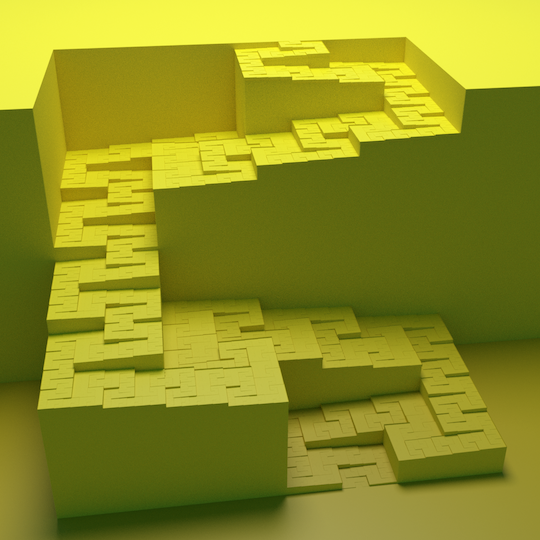}\hfill
}
\vskip0.01\hsize
\hbox to\hsize{%
\includegraphics[width=0.49\hsize]{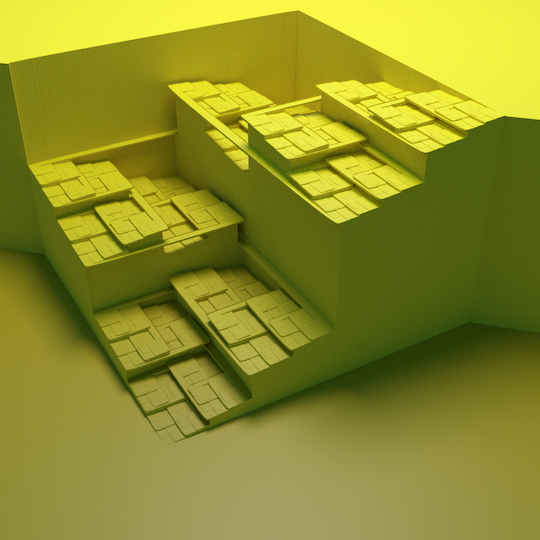}\hfill
\includegraphics[width=0.49\hsize]{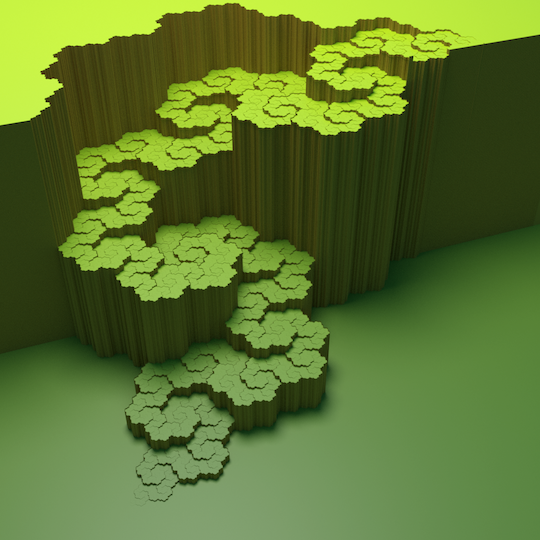}%
}
\vskip0.01\hsize
\hbox to\hsize{%
\includegraphics[width=0.32\hsize]{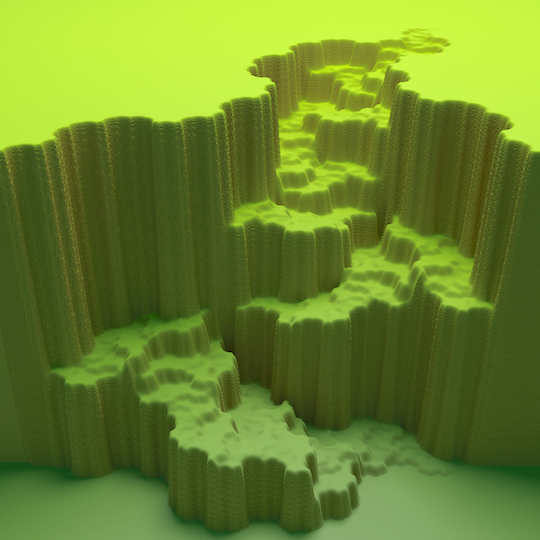}\hfill
\includegraphics[width=0.32\hsize]{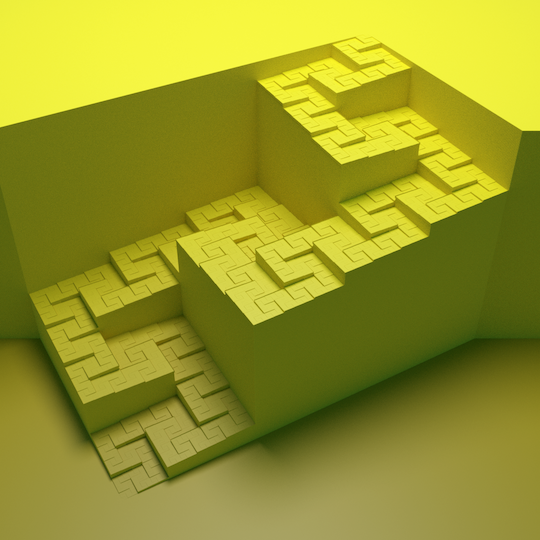}\hfill
\includegraphics[width=0.32\hsize]{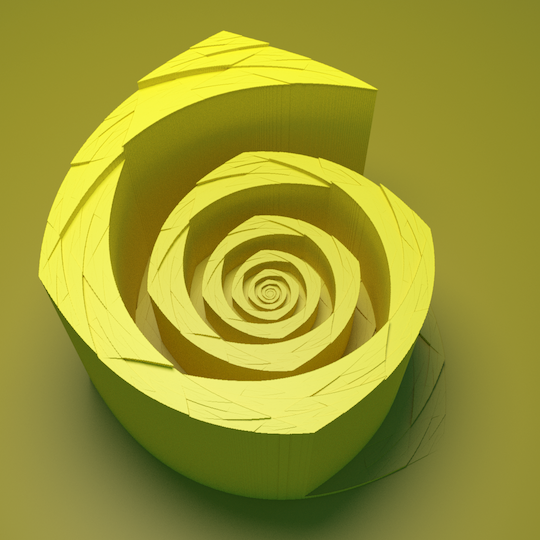}%
}
\caption{On top: 
Hilbert's curve~\cite{Hil1891} and the $\beta\Omega$~\cite{Wie02} curve from Figure~\ref{fig:hilbertsketch}, now easy to compare.
In the middle: Double-Gray-code~\cite{Fal86} and a curve filling half of a Rauzy fractal~\cite{Rau82}. Bottom: a curve from Ventrella~\cite{Ven12} (p86) filling a fractal ``pinwheel'' tile~\cite{BMT16}, rendered ``eroded'';  
the Peano curve of Figure~\ref{fig:confusingsketches}a;
and a close-up of the point at 1/3 of Pólya's curve ($\zeta = 5$).}
\label{fig:examples}
\end{figure}

Note that in these renderings, the space under the plane-filling trail is filled, thus creating the vertical walls in the landscape. Without these vertical walls, the visualisation would be rather ambiguous: it would be difficult or impossible to see if a gap between two points $p$ and $q$ in the three-dimensional view is strictly vertical, or if the gap also exists in the projection on a horizontal plane. The first would imply that $p$ and $q$ are the same point in two-dimensional space that is visited by the curve twice; the second could hint at gaps in the two-dimensional image of the curve. Filling up the space under each point of the trail eliminates the ambiguity that results from projecting the three-dimensional model onto a two-dimensional viewing plane. In many of the examples in Figure~\ref{fig:examples}, an additional visual frame of reference is provided by a background that consists of a low plane in the front, a cliff, and a high half-plane in the back. 

\begin{figure}
\hbox to\hsize{%
\includegraphics[width=0.49\hsize]{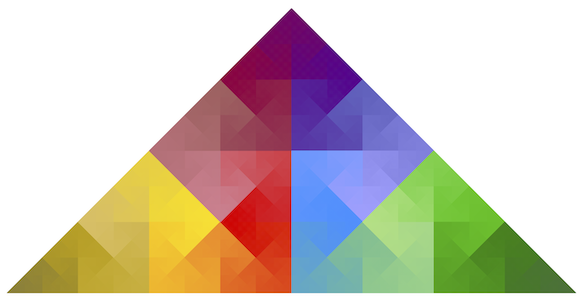}\hfill
\includegraphics[width=0.49\hsize]{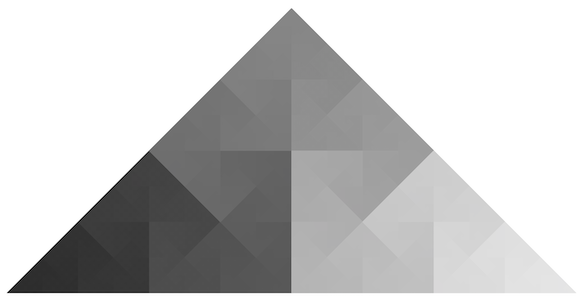}%
}
\caption{Colour and grey-scale progression images of Pólya's curve.}
\label{fig:colourgradient}
\end{figure}

\subparagraph{Alternative visualisations}
Alternative visualisation methods that come closest to meeting the same goals render the $t$-coordinate as values on a grey or colour scale instead of elevation. Indeed, such renderings are quite common, and they truly show the curve, not merely its definition, colouring the entire image according to the order in which points are visited. Figure~\ref{fig:colourgradient} shows two examples.
In comparison to the plane-filling trail method, the following differences can be observed. The colour progressions show the complete curve without distortion, whereas the plane-filling trail methods are affected by perspective distortion and occlusion. However, the colour progressions have less discernible detail. One complicating factor is that the perception of colour depends very much on the local context. As a result, it is sometimes hard to see where the curve goes: in the grey-scale image, the Pólya curve sometimes seems to alternate between becoming brighter and darker, and may even seem to cross itself. The colour image shows more detail than the grey-scale image, but has other problems. Colour gradients do not expose similarities between different parts of the curve well and they do not communicate distances along the curve well, because colour difference establishes, at best, an ordinal scale, not an interval scale.  Our visual system may not perceive translations and scalings on the colour scale as similarity transformations. This is where the plane-filling trails excel, since elevation gives us an effective interval scale on which we can read distances along the curve.

Another interesting alternative are the three-dimensional models by Irving and Segerman~\cite{SI13} that stack different refinement levels according to a definition of the curve. However, these models are hard to ``read'' when presented as a two-dimensional printed image, and they are inherently dependent on the chosen definition of the curve. That is fine if one wants to illustrate the definition, but it is a shortcoming if one wants to be able to reveal the equivalence of different definitions by producing the same image in such cases.

\subparagraph{The \pftrail\ tool}
The \pftrail\ tool reads the definition of a plane-filling traversal in the format from Ventrella~\cite{Ven12}, extended to support discontinuities and multiple refinement rules (known as \emph{generators}). Thus, the various traversals that have been proposed in the computer science literature~\cite{BuH16,Sam06,Wie02} can all be rendered and it is easy to explore new designs. Traversals are not confined to an integer grid, so we can also visualise interesting traversals related to, for example, the Rauzy fractal~\cite{Rau82} (see Figure~\ref{fig:examples}). For rendering, the traversal is sampled and drawn on a grid of hexagonal cells; thus \pftrail\ operates without any knowledge of the shape that is filled by the traversal (which can be a complicated fractal). 
The tool offers various options to control parameters such as camera position, resolution of the rendering grid, visualisation style, colour scheme, and what to do with sample points of different elevations in the same cell. Small ``parapets'' can be added at the top of high cliffs, to make the edge of the cliff more visible when seen from the ``mountain'' side, thus enhancing the perception of depth.
The output is a \collada\ file that can be rendered with, for example, Blender; if the resolution is not too high, it can also be moved around in Blender in real time. 

\subparagraph{Sampling density}
The sampling algorithm in \pftrail\ adapts automatically to the grid size. To make use of the resolution of the grid and to avoid spurious holes in the image, the sample points must be sufficiently dense, so that, at least, the following conditions are satisfied: (1) if a grid cell is partially covered by the curve, then there is at least one sample point in that grid cell or one of its neighbours; (2) if a grid cell is entirely covered by the curve, then there is at least one sample point inside that grid cell; (3) if an edge between two grid cells is entirely covered by the curve, then there is a sample point in at least one of the two cells. The implementation exploits the curve's self-similar structure to compute a good upper bound on the radius $r$ such that any section of the curve whose endpoints are a distance $d$ apart, can only visit points that are within distance $dr$ from the closest end point. To fulfil all requirements, it then suffices to make sure that the distance between sample points along the curve is at most half of the grid edge length divided by $r$.

With the current implementation, I found grid sizes of at least $2000 \times 2000$ hexagonal cells to be feasible on my laptop; the model is then generated within a few minutes. Blender could generate a poster-size image ($5000 \times 5000$ pixels) from such a model in a few hours, using the Cycles ray-tracer. For the figures in this paper I used a grid of at least $500 \times 500$ cells, from which Blender generated an image in a few minutes.

\subparagraph{Polynomial close-up}
Special features of the software include ``polynomial'' close-up: given a focus point $p$ and a zoom parameter $\zeta$, any point $q$ at distance $r$ from $p$ is moved to a point at distance $r^{1/\zeta}$ from $p$. More precisely, if we take $p$ as the origin of the $x,y,t$-coordinate system, then any point $q$ with coordinates $(x,y,t) = (r \cos\phi, r \sin\phi, t)$ is mapped to $(r^{1/\zeta} \cos\phi, r^{1/\zeta} \sin\phi, \mathrm{sign}(t) \cdot |t|^{1/(1.5\zeta-0.5)})$. This allows us to zoom in on features that remain invisible in normal close-up views. 

\begin{figure}
\centering\includegraphics{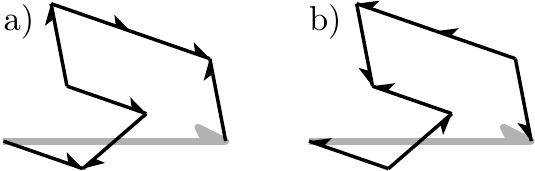}
\caption{Definitions of a) the Gosper curve~\cite{Gar76} and b) the Inner-flip~\cite{Ven12} or Alternating~\cite{Sch12} Gosper curve.}
\label{fig:gosperdefinition}
\end{figure}

\begin{figure}
\hbox to\hsize{%
\includegraphics[width=0.32\hsize]{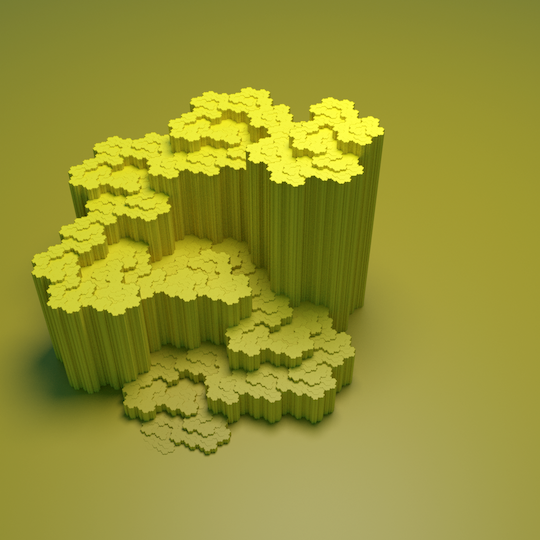}\hfill
\includegraphics[width=0.32\hsize]{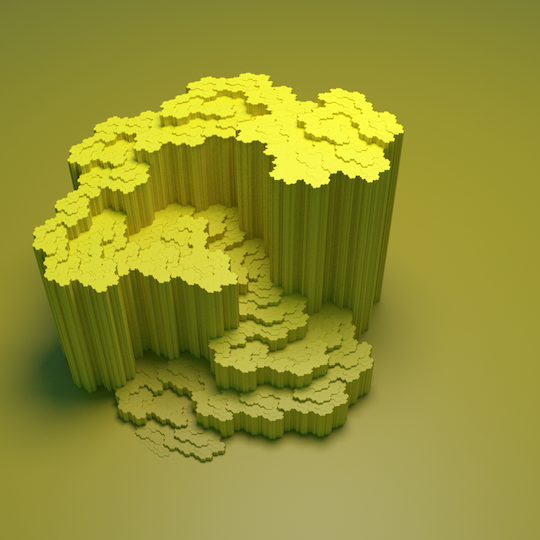}\hfill
\includegraphics[width=0.32\hsize]{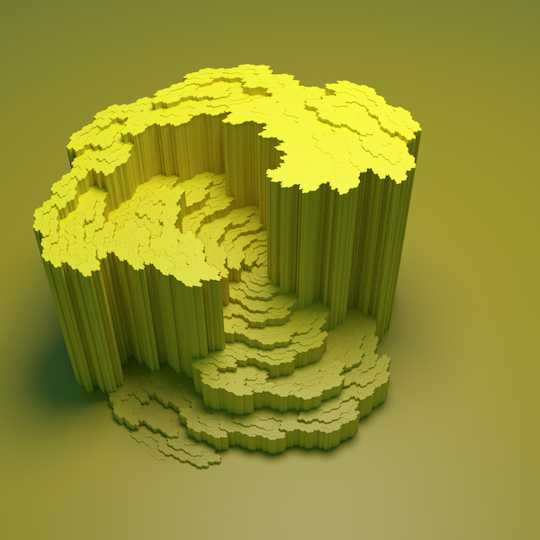}%
}
\vskip0.01\hsize
\hbox to\hsize{%
\includegraphics[width=0.32\hsize]{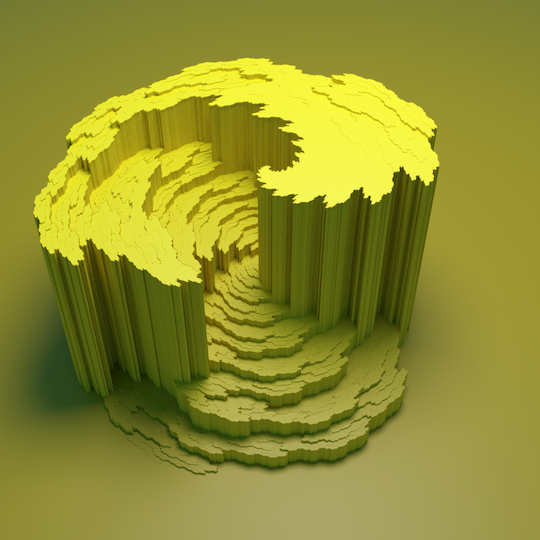}\hfill
\includegraphics[width=0.32\hsize]{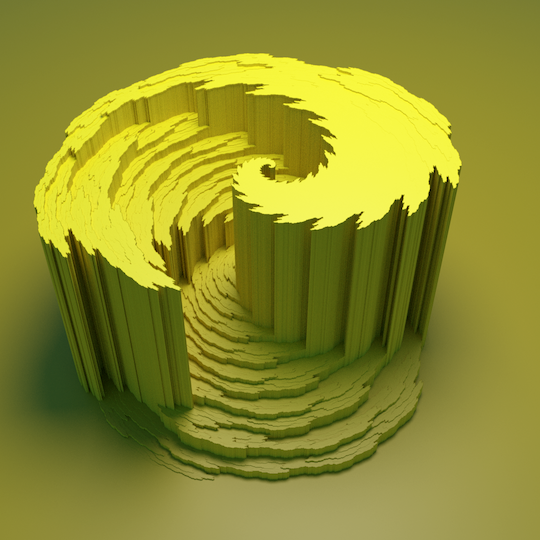}\hfill
\includegraphics[width=0.32\hsize]{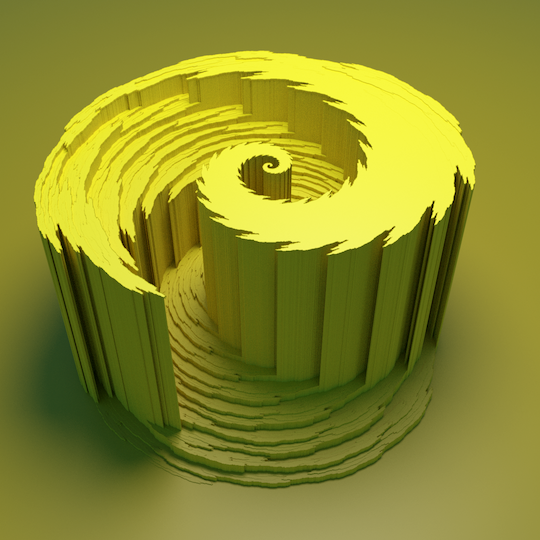}%
}
\vskip0.01\hsize
\hbox to\hsize{%
\includegraphics[width=0.32\hsize]{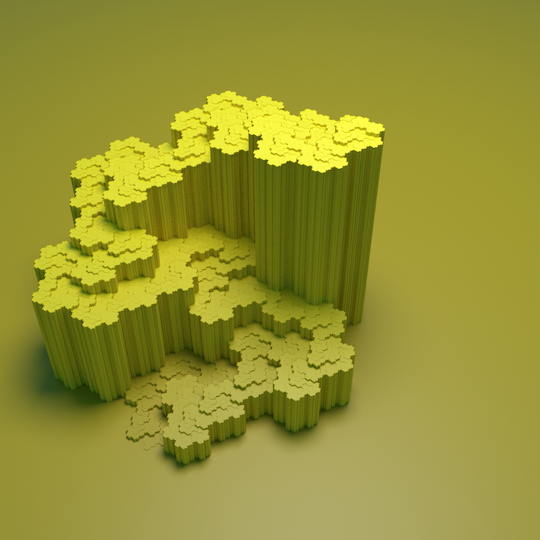}\hfill
\includegraphics[width=0.32\hsize]{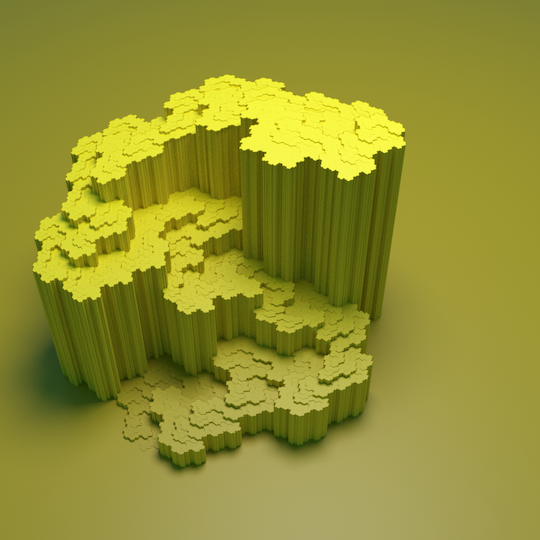}\hfill
\includegraphics[width=0.32\hsize]{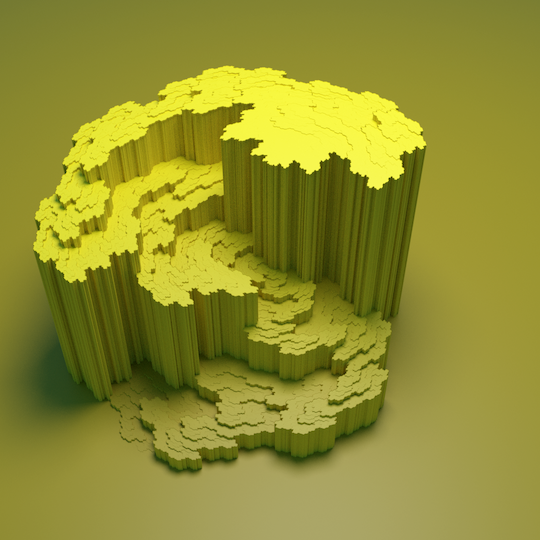}%
}
\vskip0.01\hsize
\hbox to\hsize{%
\includegraphics[width=0.32\hsize]{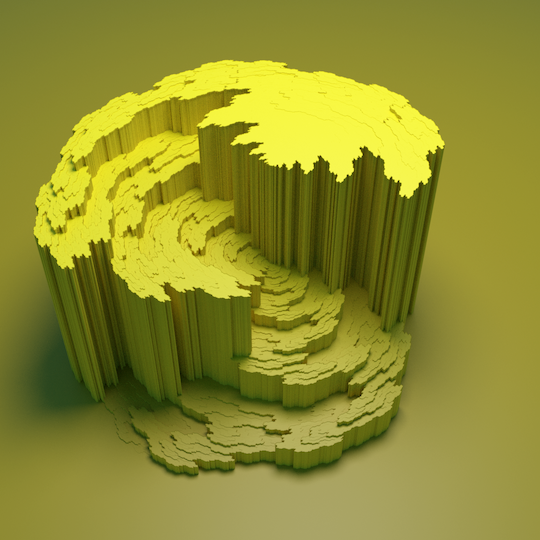}\hfill
\includegraphics[width=0.32\hsize]{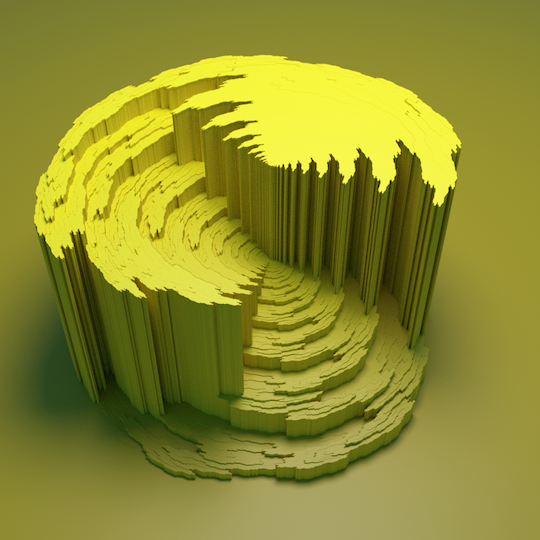}\hfill
\includegraphics[width=0.32\hsize]{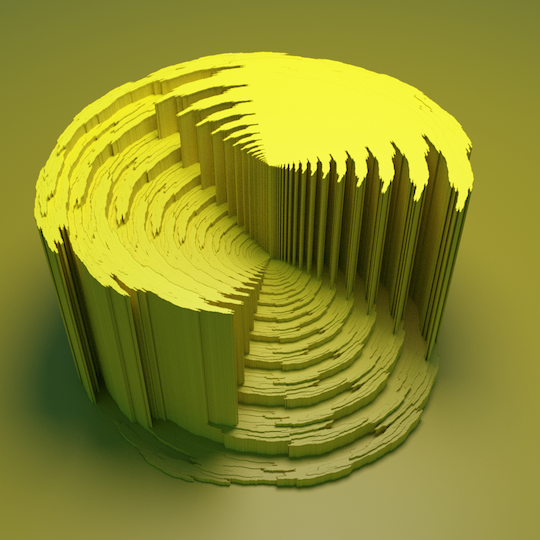}%
}
\caption{Polynomial close-up views with $\zeta = 1, 1.4, 2, 3, 5, 8$ for the Gosper curve~\cite{Gar76} (top two rows) and the Inner-flip Gosper curve~\cite{Sch12,Ven12}.}
\label{fig:gosperzoom}
\end{figure}

For example, consider the Gosper curve, which is defined in Figure~\ref{fig:gosperdefinition}a and shown as a plane-filling trail in Figure~\ref{fig:gosperzoom}, top left. The curve follows a tessellation with tiles arranged in a hexagonal grid pattern. In particular, $f(2/7)$ is the vertex where the second, third, and seventh tile meet. At such vertices, the tiles wind around each other like spirals. From geometric descriptions of the tiling~\cite{Hav11} one can derive that the spirals are logarithmic spirals with a growth rate of a factor $\sqrt{7}$ per rotation of $\arctan\sqrt{3/25}$, which is approximately $9 \cdot 10^7$ per revolution. Thus, at such a vertex, we cannot say in what direction to travel to enter a particular one of the three adjacent tiles: any ray from $f(2/7)$ with positive length, no matter the direction and no matter how short, crosses all three tiles.

In contrast, we may consider a variation of the Gosper curve in which all the defining line segments are reflected such that the arrowheads move to the other end (see Figure~\ref{fig:gosperdefinition}b). Ventrella~\cite{Ven12} calls this variation \emph{Inner-flip Gosper}. The Inner-flip Gosper does not exhibit this logarithmic spiral behaviour at the vertices of the tessellation. The boundaries between the tiles are still fractals, but they are confined to 30 degrees' wedges that meet at the vertices of the tessellation. Between these wedges, there are 90 degree's wedges that each lie entirely inside a single tile. 

No normal close-up views could show this remarkable difference in character between these two plane-filling curves, since no normal close-up view could show logarithmic spirals that shrink as fast as a factor $9 \cdot 10^7$ per revolution. But with polynomial close-up views, we can reveal these spirals and wedges clearly. Figure~\ref{fig:gosperzoom} shows the result of zooming in, with increasing values of $\zeta$, on the point $f(2/7)$ in the Gosper and Inner-flip Gosper curves.

\subparagraph{Limitations of path-filling trails}
The attentive viewer may notice a particular aspect of the plane-filling trails: the visualisation is not scale-independent. A section of the curve that is similar to the curve as a whole will look like the curve as a whole, but considerably flatter. If the curve is scaled down by a factor $s$ in the plane, its area shrinks by a factor $s^2$, and so does the one-dimensional pre-image, that is, the projection on the vertical axis. Relative to the horizontal dimensions, the vertical dimension is therefore scaled down by an additional factor $s$. This has advantages and disadvantages: on the one hand, it means that we have to make the model quite high to be able to see details; on the other hand, it may make it easier to focus on the coarser levels of detail. Despite the flattening, I believe the plane-filling trails still tend to show a wider range of levels of detail clearly than traditional drawings achieve with line work and colours, but one should be aware that the flattening effect can be a limiting factor. From a certain resolution onwards, the amount of detail that can be shown along a path-filling trail will scale linearly with the height, not the area, of the drawing.

Another aspect that one should be aware of, is that similarity by reversal, that is, by reflection of the pre-image of the curve, may not be as easy to recognize as similarity by rotation, reflection, translation and scaling of the image. It may not be visually obvious that one part of the terrain is the same as another part of the terrain turned upside down; recognizing such similarities may require that the viewer makes a conscious effort to look for them.

\subparagraph{Remaining challenges}
There are several ways in which \pftrail\ could be improved. The following issues are high on the priority list. 

By design, \pftrail\ currently operates without being given any knowledge of the image of the plane-filling traversal. Thus it is flexible, and the images that are produced depend mostly on the traversal itself, and hardly on the particular iterated function system that is used to define it. Of course the iterated function system determines the sample points that are used, but due to the conditions on the sampling density as explained above, this effect is small. However, there are circumstances in which the current approach does not suffice to produce the perfect figure, and in which it would be useful to be able to give \pftrail\ some hints. 

In particular, the following circumstances can cause suboptimal rendering. First, in some traversals, jumps create bridges and tunnels that start or end on gentle slopes. These bridges and tunnels interfere with the relief on that slope, so that relief, tunnel entrances, or bridgeheads are distorted or disappear entirely. Although \pftrail\ currently provides several options to control what happens in such cases, it seems that in some cases, the best solution would be a context-dependent asymmetric rendering of the bridge or tunnel. Second, in many curves, there are narrow passages where the trail connects two regions that only meet in a point. For clarity, it would often be good to make such passages wider---but to be able to do so, \pftrail\ would need to know where the narrow passages are. Third, interference between the rendering grid pattern and the sampling pattern may lead to small but obtrusive visual artefacts. Oversampling, which is an option in \pftrail\ already, may reduce or eliminate these artefacts, but this solution can be quite inefficient. Given a simple description of the image of each curve section, it would be possible to guarantee that sample points are placed in \emph{all} grid cells intersected by the traversal. This would eliminate artefacts that stem from the sampling pattern.

Finally, to enable the production of 3D-printable models, a solution must be implemented that prevents or eliminates (near-)degeneracies around bridges and saddle points, without interfering with the semantics of the model.

\bibliography{plane-filling-trails}

\end{document}